\documentclass[a4paper,11pt]{article}
\usepackage{jheppubcustom}
\usepackage[T1]{fontenc}
\usepackage[utf8]{inputenc}
\usepackage{braket}
\usepackage{xcolor}
\usepackage{subcaption}

\graphicspath{ {./plots/} }

\def\dB{\text{d}_\text{B}}

\def\F{\text{Fid}}
\def\rhoO{\rho_\Omega}
\def\rhoD{\rho_D}
\def\rhoC{\rho_C}
\def\Ctop{C_\text{top}}
\def\HSA{H_{\mathcal{S}\mathcal{A}}}

\def\HA{H_\mathcal{A}}
\def\HE{H_\mathcal{E}}
\def\ketOmegaS{\ket{\Omega}_\mathcal{S}}
\def\ketCS{\ket{C}_\mathcal{S}}

\def\Tr{\textsf{Tr}}

\DeclareSymbolFont{bbold}{U}{bbold}{m}{n}
\DeclareSymbolFontAlphabet{\mathbbold}{bbold}
\newcommand{\1}{\mathbbold{1}}

\title{\boldmath  Decoherence: A Numerical Study}

\author{$\text{Chris Nagele}^{\dagger}$,}
\author{$\text{Oliver Janssen}^{\ddagger,*,\star} \text{ and}$}
\author{$\text{Matthew Kleban}^{\ddagger}$}
\affiliation{${}^\dagger$ Department of Astronomy, Graduate School of Science, the University of Tokyo, Tokyo, 113-0033, Japan}
\affiliation{${}^\ddagger$ Center for Cosmology and Particle Physics, New York University, 726 Broadway, New York, NY 10003, USA}
\affiliation{${}^{*}$ International Centre for Theoretical Physics, Strada Costiera 11, 34151 Trieste, Italy}
\affiliation{${}^{\star}$ Institute for Fundamental Physics of the Universe, Via Beirut 2, 34014, Trieste, Italy}
\emailAdd{christophernagele@gmail.com}
\emailAdd{oliverjanssen@nyu.edu}
\emailAdd{kleban@nyu.edu}
\abstract{
We study quantum decoherence numerically in a system consisting of a relativistic quantum field theory coupled to a measuring device that is itself coupled to an environment. The measuring device and environment are treated as quantum, non-relativistic particles. We solve the Schr\"odinger equation for the wave function of this tripartite system using exact diagonalization. Although computational limitations on the size of the Hilbert space prevent us from exploring the regime where the device and environment consist of a truly macroscopic number of degrees of freedom, we  nevertheless see clear evidence of decoherence: after tracing out the environment, the density matrix describing the system and measuring device evolves quickly towards a matrix that is close to diagonal in a subspace of pointer states.  
}

\begin{document} 
\maketitle
\flushbottom

%%%%%%%%%%%%%%%%%%%%%%%%
\section{Introduction}
%%%%%%%%%%%%%%%%%%%%%%%%
Quantum mechanics is the most successful theory in the history of science.  It explains phenomena as disparate as fluorescent lights, nuclear reactions and the origin of structure in the universe, and is precise enough to calculate the magnetic moment of the electron to 12 significant figures.  Despite these extraordinary successes, certain fundamental features of the theory remain obscure.  Chief among them is the question of how quantum mechanics relates to the classical world -- how (or if) the Born rule for calculating the probabilities of measurement results from the quantum wave function should be understood, why classical (rather than quantum) physics accurately describes the macroscopic world and how to understand the apparent collapse of the wave function following a measurement.

A widely-accepted idea that bears on these mysteries is decoherence \cite{Zeh:1970zz,Zeh:1973,PhysRevD.24.1516,PhysRevD.26.1862,Joos:1984uk} (for reviews see \cite{RevModPhys.75.715,Schlosshauer:2003zy,Schlosshauer:2019ewh}). The idea is that the interaction of a microscopic quantum system with a macroscopic measuring device (or ``apparatus'') and an environment should, via  Schr\"odinger time evolution, cause the  wave function to evolve towards a specific form.  Suppose the initial  quantum state of the system was a superposition of eigenstates of the operator being measured that is unentangled with the environment and apparatus. The measurement process should generate entanglement such that after tracing out the environment, the final wave function will approximate a mixed state that is a sum over eigenstates times appropriate states of the apparatus, weighted by the Born-rule probabilities.  For all observables restricted to the system-apparatus subspace, this mixed state gives predictions that are precisely identical to a \emph{classical} probabilistic mixture of these pure quantum eigenstates with the Born-rule probabilities (see \S \ref{backgroundsec}) -- just what traditional interpretations predict following the interaction of the system with a measuring device \cite{dirac1930book}.  Hence, if this  evolution takes place it can be regarded as at least a partial explanation for wave function collapse with the correct probabilities.  Unfortunately, the difficulties inherent in solving for the  time evolution of the quantum state of macroscopic systems make it very difficult to establish whether this is the case.

In this paper we numerically evolve the wave function for a quantum system that we consider to be a  model of a realistic experiment. The system under study is an interacting, relativistic quantum field theory (QFT) -- the massive Schwinger model, or quantum electrodynamics defined in one space and one time dimension. The QFT is coupled to a measuring device or ``apparatus'' (modeled as a heavy quantum particle) via a von Neumann-type interaction. The apparatus reacts to the presence of charges at a particular location in the Schwinger system. The apparatus is in turn coupled to an environment. We take the environment to be a light particle that interacts with the apparatus via a local (in position) interaction potential. We think of this light particle as a molecule of gas that the experimenter failed to evacuate from the tube containing the apparatus. After solving the Schr\"odinger equation for the full tripartite system, we trace over the environment to produce a density matrix for the system plus apparatus, and compare the resulting mixed state to the one predicted by decoherence. The distance between these states is a quantitative measure of the degree to which the system has decohered.

The advantage of our approach is that we make no approximations apart from numerical discretization, nor do we assume anything about the interpretation of quantum mechanics or the Born rule. Instead, we simply evolve the wave function and compare it to the one predicted by the theory of decoherence.  Furthermore (in contrast to the  toy examples usually studied in this context) the system being measured is a causal, relativistic QFT that is similar in many ways to the QFTs believed to describe the fundamental physics of our world.  This opens the door to investigating \emph{causal} aspects of decoherence, such as how localized measures of decoherence spread in spacetime.

%%%%%%%%%%%%%%%%%%%%%%%%
\subsection{Background and previous work} \label{backgroundsec}
%%%%%%%%%%%%%%%%%%%%%%%%
The basic formalism of decoherence (see e.g. \cite{Schlosshauer:2003zy}) takes place in a tripartite Hilbert space $\mathcal{H}$ consisting of the tensor product of the subsystems $\mathcal{S}$ (the system), $\mathcal{A}$ (the apparatus) and $\mathcal{E}$ (the environment): $\mathcal{H} = \mathcal{H}_\mathcal{S} \otimes \mathcal{H}_\mathcal{A} \otimes \mathcal{H}_\mathcal{E}$. An idealized version of decoherence would be a state that evolves in the following way under the Schr\"odinger equation as the system interacts with the apparatus:
\begin{align}
    \ket{\psi(0)} = \left( \sum_n c_n \ket{s_n} \right) \ket{a_0} \ket{e_0} &\overset{(1)}{\longrightarrow} \left( \sum_n c_n \ket{s_n} \ket{a_n(t_1)} \right) \ket{e_0} \\
    &\overset{(2)}{\longrightarrow} \left( \sum_n c_n \ket{s_n} \ket{a_n(t)} \ket{e_n(t)} \right) = \ket{\psi(t)} \,, \label{psitassump}
\end{align}
where $0 < t_1 < t$.  
By definition, we do not measure properties of the environment and so it should be traced out. This leads to the following reduced density matrix for the $\mathcal{S}\mathcal{A}$ subsystem:
\begin{equation} \label{rhoSAexact}
    \rho_{\mathcal{S}\mathcal{A}}(t) = \textsf{Tr}_\mathcal{E} \ket{\psi(t)}\bra{\psi(t)} = \sum_{n,m} c_n c_m^* \braket{e_m(t)|e_n(t)} \, \ket{s_n} \ket{a_n(t)} \bra{s_m} \bra{a_m(t)} \,.
\end{equation}
A key mathematical claim of decoherence is that the environment states $\{ \ket{e_n(t)} \}$ will  rapidly  become orthogonal to each other so that $\braket{e_m(t)|e_n(t)} \rightarrow \delta_{m,n}$. Then 
\begin{equation} \label{rhoSA}
    \rho_{\mathcal{S}\mathcal{A}}(t) \rightarrow \sum_n |c_n|^2 \ket{s_n} \ket{a_n(t)} \bra{s_n} \bra{a_n(t)} \,.
\end{equation}
The collection of states $\{ \ket{s_n} \ket{a_n(t)} \} \subset \mathcal{H}_\mathcal{S} \otimes \mathcal{H}_\mathcal{A}$ are called ``pointer states'': they retain their correlation despite an interaction with the environment. Equation \eqref{rhoSA} indicates that $\rho_{\mathcal{S}\mathcal{A}}(t)$ becomes diagonal in (a subset of) pointer states.\footnote{The pointer states do not form a basis -- generally they form an overcomplete set. E.g. in the case of a harmonic oscillator $\mathcal{S}$ coupled to a heat bath $\mathcal{E}$, the pointer states are the coherent states of the harmonic oscillator, provided the friction constant of the environment is much smaller than the oscillator's frequency \cite{Paraoanu_1999}.} The interaction with the environment has suppressed interference terms of the type $\ket{s_n} \ket{a_n(t)} \bra{s_m} \bra{a_m(t)}, m \neq n$ that would have been present in the $\mathcal{S}\mathcal{A}$ density matrix had $\mathcal{A}$ not been coupled to $\mathcal{E}$.

However, in realistic systems all states  become entangled with the environment at some level.  This means no state will remain pure after tracing over $\mathcal{E}$.  Given this, one way to characterize pointer states is the ``predictability sieve'' \cite{Zurek:1994zq}.  Given an initial state $\ket{\chi_0} = \ket{s} \ket{a} \in \mathcal{H}_\mathcal{S} \otimes \mathcal{H}_\mathcal{A}$ and a typical environment state $\ket{e_r}$, we can consider the entropy of the reduced density matrix $\rho_{\mathcal{S}\mathcal{A}}(t) = \textsf{Tr}_\mathcal{E} \ket{\psi(t)} \bra{\psi(t)}, S(t) = - \textsf{Tr}_{\mathcal{S}\mathcal{A}} \left( \rho_{\mathcal{S}\mathcal{A}} \, \textsf{log} \, \rho_{\mathcal{S}\mathcal{A}} \right)$, where $\ket{\psi(t)}$ is the time evolution of the initial state $\ket{\chi_0} \ket{e_r}$.  Pointer states are those $\ket{\chi_0}$ for which $S(t)$ rises slowly on timescales typical of the dynamics of $\mathcal{S}$ and $\mathcal{A}$.

Experiments conducted on the system and apparatus only cannot distinguish the mixed state  \eqref{rhoSA}  from a classical statistical ensemble of quantum states $\{ \ket{s_n} \ket{a_n(t)} \}$ with probabilities $|c_n|^2$ \cite{preskill}.  To see this, 
note that  for any observable $\mathcal{O}$, using \eqref{rhoSA} we have
\begin{equation} \label{Oexp}
    \langle \mathcal{O} \rangle = \textsf{Tr}_{\mathcal{S}\mathcal{A}} \left[ \mathcal{O} \rho_{\mathcal{S}\mathcal{A}}(t) \right] = \sum_n |c_n|^2 \bra{s_n} \bra{a_n(t)} \mathcal{O} \ket{s_n} \ket{a_n(t)}  \,,
\end{equation}
which is manifestly identical to $\langle \mathcal{O} \rangle$ computed in the classical ensemble of quantum states mentioned above.  This holds regardless of whether the states $\ket{s_n} \ket{a_n(t)}$ are orthogonal.  Similarly, the convex combination of \emph{mixed} states
\begin{equation} \label{rhoSAden}
    \rhoD(t) = \sum_n |c_n|^2 \rho_n(t) \,
\end{equation}
is  equivalent in the same sense to a classical statistical ensemble of mixed states $\rho_n$ with probabilities  $|c_n|^2 $.  If the $\rho_n(t)$ are time-evolved pointer states (which, as just mentioned, will not in general remain pure), then  $\rhoD(t)$ can be thought of as the state predicted by decoherence: a classical statistical ensemble of time-evolved pointer states.

Despite the crucial role attributed to decoherence in understanding the relation of quantum mechanics to the classical world (see e.g. \cite{decohbook96,openquantumbook,Joos:1984uk}), analytical studies of it have been limited to specific interactions and toy models (see e.g. \cite{Albrecht:1992rs,Allahverdyan:2011cx,Nieuwenhuizen:2014nea}, and the review \cite{Schlosshauer:2019ewh} and references therein), while numerical studies are limited by the computational cost of simulating exponentially large Hilbert spaces.  One  numerical study \cite{adami_negulescu_2012} considers two non-relativistic particles interacting on an interval. The heavier of the two particles is taken as $\mathcal{S}$ while the lighter is $\mathcal{E}$; there is no $\mathcal{A}$. 
The authors consider the time evolution of an initial product state consisting of two Gaussian lumps in position space for the heavy particle that move towards each other, times a single Gaussian lump for the lighter particle. With no interaction, the position space probability density of the heavier particle $\mathcal{S}$ (obtained by tracing out the lighter particle $\mathcal{E}$) develops an interference pattern as the two lumps approach one another and  overlap. With an appropriate interaction between the particles this inference pattern is destroyed: the  heavy particle position space density is well-approximated by the sum of the two respective heavy particle densities even after they overlap. This is an example of the general decoherence mechanism described above applied to the specific observable $\mathcal{O} = x_\textsf{heavy}$ in Eq. \eqref{Oexp}, with the pointer states being the independent free time evolutions of the two heavy particle lumps.

In this work, we make no approximations (apart from those inherent in discretizing the system) and as a result any pointer state will become entangled with the environment to some extent -- but more slowly than a typical state would (this is the predictability sieve criterion for pointer states described above).  As a result, the sum over pure states in \eqref{Oexp} must be replaced by the corresponding convex combination of density matrices \eqref{rhoSAden} where the $\rho_n(t) \equiv \rho_{\mathcal{S}\mathcal{A}}(t)$ for $c_n = 1, c_{m \neq n} = 0$ (in other words, $\rho_n(t)$ is the density matrix resulting from tracing over the environment when the initial state is the single pointer state $\ket{s_n} \ket{a_n(t)}$).  In the rest of this work, our primary measure of decoherence will be the distance between the density matrix $\rhoD(t)$ defined in \eqref{rhoSAden} and the exact mixed state $\rho_{\mathcal{S}\mathcal{A}}(t)$ defined in \eqref{rhoSAexact}.

%%%%%%%%%%%%%%%%%%%%%%%%
\section{Numerical methods and system}
%%%%%%%%%%%%%%%%%%%%%%%%
The quantum theory we study consists of three parts:  an interacting relativistic quantum field theory in one spatial dimension ($\mathcal{S}$) -- the massive Schwinger model  (quantum electrodynamics in 1+1) -- coupled via a von Neumann \cite{VN1932} type interaction to a massive, non-relativistic particle (the measuring device or apparatus $\mathcal{A}$) that is in turn coupled to a much lighter non-relativistic particle (the environment $\mathcal{E}$) via a local (in position) interaction. We regard this light particle as an air molecule that our experimenter has failed to evacuate from the apparatus' cavity. This is the classic tripartite system-apparatus-environment theory  considered in discussions of decoherence and reviewed in \S\ref{backgroundsec}.

The Schwinger model Hamiltonian is discretized on a staggered lattice with spacing $a$, with electrons on even sites and positrons on odd sites. Using a Jordan-Wigner transformation, the discretized fermionic Hamiltonian is mapped to a spin Hamiltonian \cite{PhysRevD.13.1043}. For more details please refer to \cite{Nagele:2018egu}.

\begin{figure}
    \centering
    \includegraphics[height = 8cm]{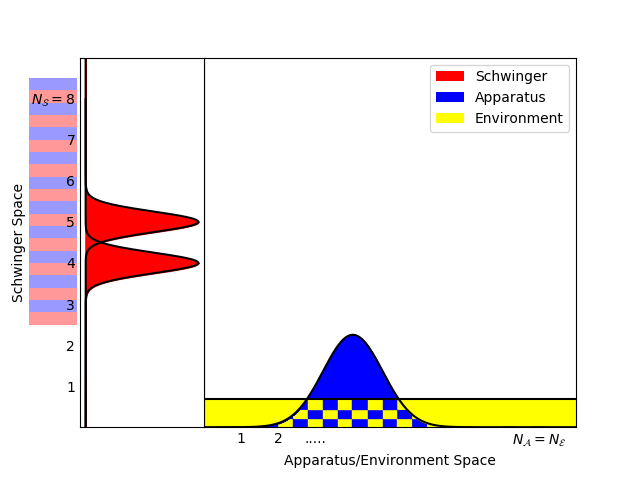}
    \caption{Schematic showing the massive Schwinger model quantum field theory (vertical axis) and the charge density for a particle-anti particle pair (red).  The horizontal axis shows the position-space density of the apparatus (heavy particle, blue) and environment (light particle, yellow).  The apparatus is coupled to the charge density of the Schwinger system at all but the bottom two sites $1$ and $2$, and the apparatus and environment are coupled via a local (in position) interaction potential (cross-hatching indicates where the initial state wave functions overlap). }
    \label{fig:schem}
\end{figure}

The Hamiltonians for the apparatus and environment are  those of massive free non-relativistic particles:
\begin{equation}
    \HA = \frac{ p^2_\mathcal{A}}{2m_\mathcal{A}}, \;\;\;\HE =  \frac{ p^2_\mathcal{E}}{2m_\mathcal{E}}
\end{equation}
where $p$ is the momentum operator and $m$ is the mass. 
The interaction Hamiltonian between the Schwinger system and the apparatus takes the form 
\begin{equation} \label{pointsys}
    \HSA = g_{\mathcal{S}\mathcal{A}}\;  \bigg[ \big(\Ctop - \bra{\Omega}_\mathcal{S} \Ctop \ketOmegaS \1_\mathcal{S} \big) \; \otimes \; p_\mathcal{A} \bigg]
\end{equation}
where $g_{\mathcal{S}\mathcal{A}}$ is a (positive) coupling constant, $\ketOmegaS$ is the ground state of the Schwinger Hamiltonian, and $\Ctop$ is the fermion density averaged over all but the bottom two Schwinger lattice sites. This interaction will cause the apparatus to move to the right in Figure \ref{fig:schem} (towards larger values of the apparatus position $x_\mathcal{A}$) when charges are present at any site except the bottom two of the Schwinger system.  We subtract $\bra{\Omega}_\mathcal{S} \Ctop \ketOmegaS$ so that the apparatus is calibrated to react as little as possible in the ground state.

Lastly, the interaction between the apparatus and the environment is of the form
\begin{equation}
    H_{\mathcal{A}\mathcal{E}} = g_{\mathcal{A}\mathcal{E}}\; V(x_\mathcal{A} \otimes \1_\mathcal{E} - \1_\mathcal{A} \otimes x_\mathcal{E})
\end{equation}
where $g_{\mathcal{A}\mathcal{E}}$ is the coupling, $x_\mathcal{A}$, $x_\mathcal{E}$ are the position operators for the apparatus and environment particles, and $V$ is  the interaction potential that we take to be Gaussian,
\begin{equation}
    V(x) = \frac{1}{\sqrt{2 \pi} \sigma} \exp \left( \frac{-x^2}{2 \sigma ^2} \right) \,,
\end{equation}
where $\sigma$ is the range of the interaction.

For simplicity and because the interaction is local in position, the number of lattice sites for the apparatus and environment are equal,  $N_\mathcal{A} = N_\mathcal{E}$. The full Hamiltonian reads
\begin{equation}
    H = \left( H_\mathcal{S} \otimes \1_\mathcal{A} + H_\mathcal{SA} \right) \otimes \1_\mathcal{E} + \1_\mathcal{S} \otimes \left( \HA \otimes \1_\mathcal{E} + \1_\mathcal{A} \otimes \HE + H_\mathcal{AE} \right) \,.
\end{equation}

%%%%%%%%%%%%%%%%%%%%%%%%
\subsection{Pointer states and numerical measures of decoherence}
%%%%%%%%%%%%%%%%%%%%%%%%
As mentioned above, decoherence is defined by the time evolution of the density matrix describing the system or the system and apparatus, after tracing over the environment.  It refers to the tendency of the density matrix to become diagonal in a special basis of so-called ``pointer states'' that depend on the details of the system and its interactions with the environment.

Given the form of the interactions, we expect the pointer states to be localized in the position basis for the apparatus, and to be (close to) charge eigenstates for the Schwinger system.  The wave function of true position eigenstates would spread out more rapidly than the characteristic timescales of the system, so we will consider   Gaussian wave functions with a standard deviation that is small enough to distinguish between apparatus locations before and after a measurement, but large enough to prevent the wave function from immediately spreading. 

Therefore we will consider an initial state of the form
 \begin{equation} \label{initialstate}
     \ket{\psi(t=0)} =2^{-1/2}( \ketOmegaS+\ketCS) \; \otimes\; \ket{0}_\mathcal{A} \; \otimes\; \ket{e}_\mathcal{E}
 \end{equation}
where $\ketOmegaS$ is the Schwinger ground state, $\ketCS$ is the Schwinger state with charges at sites 1 and 2 (for details see \cite{Nagele:2018egu}), $\ket{0}_\mathcal{A}$ is a Gaussian in position for the apparatus and we choose $\ket{e}_\mathcal{E}$ to be an environment state that is completely de-localized in position space  (so that the air molecule is equally likely to be found anywhere on the interval).  For simplicity we focus  just on these two pointer states and their equally weighted linear combination \eqref{initialstate}.

\begin{figure}
    \centering
    \includegraphics[height = 13cm]{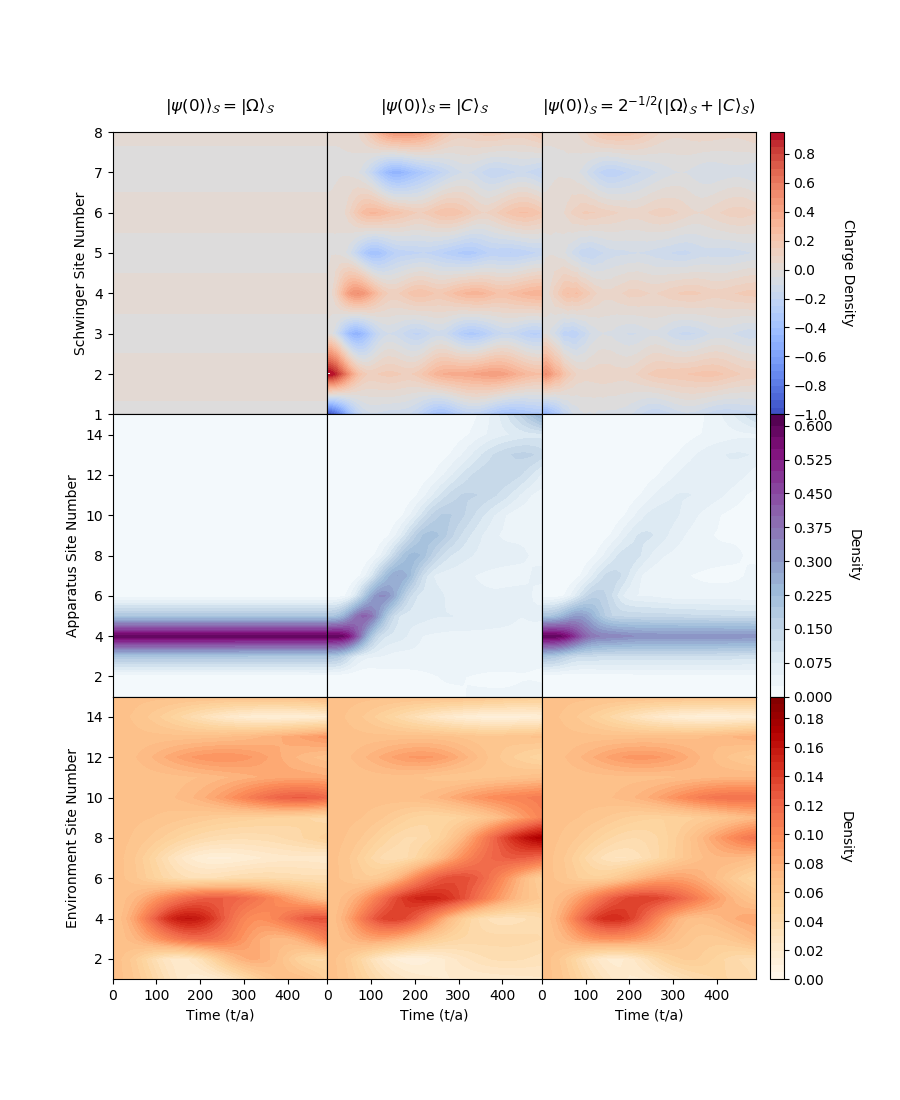}
    \caption{The expectation value of the charge density for the Schwinger model, the position of the apparatus, and the position of the environment (top, middle, and bottom rows) for three initial Schwinger states $\ketOmegaS, \ketCS$, and $2^{-1/2}\left( \ketOmegaS + \ketCS \right)$ (cf. (\ref{initialstate})) (left, middle, and right columns).}
    \label{fig:density}
\end{figure}

Due to the boundary conditions the charges in the Schwinger state $\ketCS$ have an upward momentum initially, so after some time they move out of the bottom two sites and enter the region where the coupling to the apparatus is active.\footnote{We consider a state with two equal and opposite charges to avoid a background electric field that would exert a net force on the charges.}
In Figure \ref{fig:density} we plot the evolution of the charge density and positions of the apparatus and environment.

We define $\rho(t) = \rho_{\mathcal{S}\mathcal{A}}(t)$ to be the mixed state resulting from the partial trace over $\mathcal{E}$ of this initial state:
\begin{equation} \label{tracedstate}
    \rho(t) \equiv \Tr_\mathcal{E} \ket{\psi(t)}\bra{\psi(t)},
\end{equation}
and $\rhoO(t), \rhoC(t)$ to be the mixed states when we choose the initial state of the system to be $\ketOmegaS \otimes \ket{0}_\mathcal{A} \otimes \ket{e}_\mathcal{E}, \ketCS  \otimes \ket{0}_\mathcal{A} \otimes \ket{e}_\mathcal{E}$ respectively and trace over $\mathcal{E}$ at time $t$.

\begin{figure}
    \centering
    \includegraphics[height = 8cm]{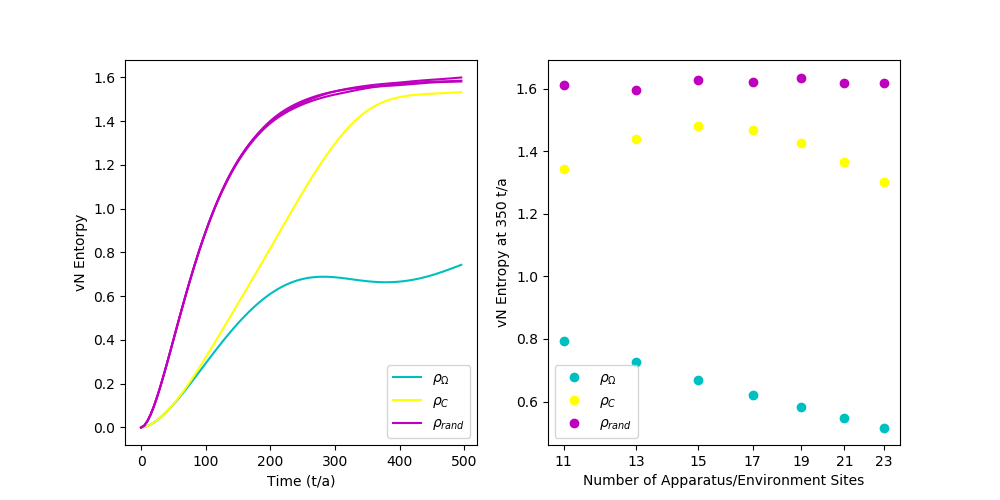}
    \caption{Left panel: time evolution of the von Neumann entropy of the mixed system+apparatus states $\rho_\Omega(t)$ and $\rho_C(t)$ obtained by the tracing over the environment of two putative initial pointer states $\ketOmegaS \otimes \ket{0}_\mathcal{A}$ and $\ketCS \otimes \ket{0}_\mathcal{A}$ (see \eqref{initialstate} and \eqref{tracedstate}), compared to several randomly chosen states. Right panel: numerical stability of the left panel when increasing the size of the apparatus/environment Hilbert space.}
    \label{fig:pointer}
\end{figure}

As described above, the predictability sieve criterion for identifying pointer states quantitatively relies on the behavior of the  entropy of entanglement with the environment. In a decohering system, the  entropy of a pointer state should remain small (relative to a random state) for some time. To check whether we have indeed identified some pointer states correctly, we plot the von Neumann entanglement entropy of two such states vs. time and compare it to the entropy of a random state (see Figure \ref{fig:pointer}).  The entropy indeed grows  more slowly for our putative pointer states than for a random state.

As mentioned previously, our quantitative measure of decoherence will be the distance between the density matrix $\rho(t)$ (cf. \eqref{tracedstate}) and 
\begin{equation} \label{decostate}
\rhoD(t) \equiv \frac{1}{2} \rhoO(t) + \frac{1}{2} \rhoC(t) .
\end{equation}
If $\rho$ and $\rho_D$ are identical this would correspond to perfect decoherence, because it would mean that the system has evolved to a state that can be interpreted as a classical statistical ensemble of the two states it would have evolved to in each pointer state separately.  Conversely, if they remain nearly as far apart as they are initially before the interaction, it would indicate that the system is not decohering. \\

\noindent There are several inequivalent definitions of distance between density matrices.  We will use the Bures distance \cite{burespaper,UHLMANN1976273}:
\begin{equation} \label{bures}
	\dB(\rho_1,\rho_2) \equiv \sqrt{1 - \sqrt{\F(\rho_1,\rho_2)}} \equiv \sqrt{1-\Tr \sqrt{\sqrt{\rho_1} \, \rho_2 \sqrt{\rho_1}}} \,.
\end{equation}
When $\rho_1, \rho_2$ are pure states $\ket{\phi_1}\bra{\phi_1}, \ket{\phi_2}\bra{\phi_2}$ the Bures distance reduces to the Fubini-Study distance $\sqrt{1-|\langle \phi_1,\phi_2 \rangle|}$. In general \cite{UHLMANN1976273,fidelity1994}
 \begin{equation*}
 	\F(\rho_1,\rho_2) = \max_{\ket{\psi_1},\ket{\psi_2}} |\braket{\psi_1,\psi_2}|^2 \,,
 \end{equation*}
 where the maximum is taken over the set of all (independent) purifications $\ket{\psi_{1,2}}$ of $\rho_{1,2}$. The Bures distance is natural here since it involves the notion of purification, where density matrices are viewed as pure states of a larger Hilbert space. 
Note that we have normalized the Bures distance so that the maximal distance between two density matrices is $\dB = 1$ when the two density matrices have support on orthogonal subspaces, and $\dB = 0$ iff.~$\rho_1 = \rho_2$.

%%%%%%%%%%%%%%%%%%%%%%%%
\section{Results}
%%%%%%%%%%%%%%%%%%%%%%%%
As we will now illustrate, the exact Schr\"odinger time evolution of  the tripartite Schwinger-apparatus-environment  quantum system indeed exhibits decoherence.  Specifically, the distance $\dB(\rho(t),\rhoD(t))$ between the mixed state $\rho$ obtained from tracing over the environment and the ideal decohered mixed state $\rhoD$ evolves from its initial value to a substantially smaller value (we believe this minimum value would be even smaller if we were to increase the size of the Hilbert space).  As expected, this decrease begins when the system starts to interact with the apparatus (and the apparatus in turn interacts with the environment).

\begin{figure}
    \centering
    \includegraphics[height = 8cm]{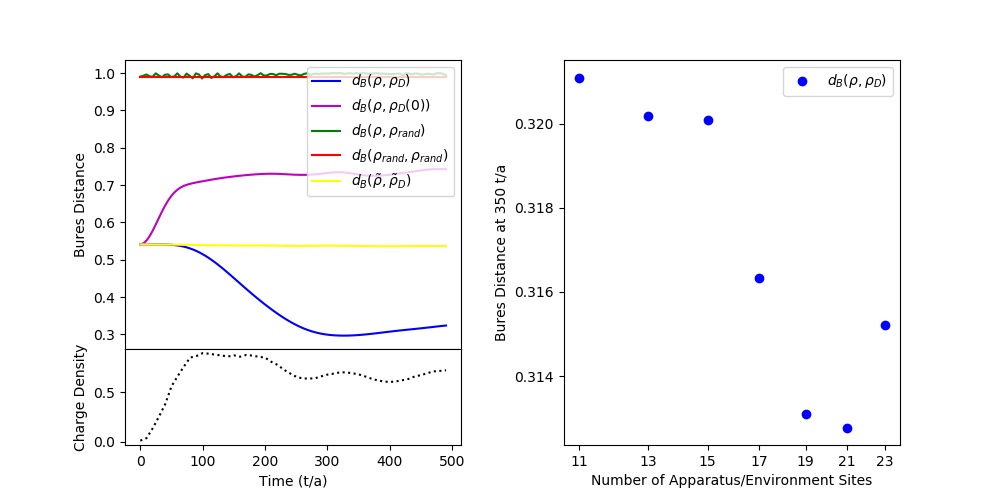}
    \caption{Left panel: Bures distance between the density matrix $\rho(t)$ obtained by tracing over the environment \eqref{tracedstate} and the fully decohered mixed state $\rhoD(t)$ \eqref{decostate} (blue). For comparison we  plot the distance between $\rho(t)$ and a random state (green), the distance between two random states (red), the distance between $\rho(t)$ and $\rho(0)$ (magenta), and the distance between $\tilde{\rho}(t)$ and $\tilde{\rho}_D(t)$ (yellow), where $\tilde{\rho}, \tilde{\rho}_D$ are defined in the same way as $\rho, \rho_D$ but relative to randomly chosen system-apparatus states (rather than pointer states).  Left lower panel: average charge density of the top six Schwinger sites. Right panel:  Bures distance $\dB(\rho, \rho_D)$ at a specific time ($t/a=350$) as a function of the size of the apparatus and environment Hilbert spaces.}
    \label{fig:decohere}
\end{figure}

As a check, we also plot the distance between $\rho(t)$ and $\rho(0)$, between the analog of $\rho(t)$ and $\rhoD(t)$ where we replace the pointer states $\ketOmegaS \otimes \ket{0}_\mathcal{A}$ and $\ketCS  \otimes \ket{0}_\mathcal{A}$ with random states, the distance between $\rho(t)$ and a random density matrix, and the distance between two random density matrices.\footnote{A random density matrix in $\mathcal{H}_\mathcal{S} \otimes \mathcal{H}_\mathcal{A}$ is chosen by acting with a random unitary (chosen according to the Haar measure) on a reference state in $\mathcal{H}_\mathcal{S} \otimes \mathcal{H}_\mathcal{A} \otimes \mathcal{H}_\mathcal{E}$ and tracing over $\mathcal{H}_\mathcal{E}$.} None of these distances decrease with time, which establishes that the behavior we are observing is indeed consistent with the predictions of decoherence. These behaviors are illustrated in Figures \ref{fig:density} and \ref{fig:decohere}.

While we do not illustrate it in a figure, setting the apparatus-environment coupling $g_{\mathcal{A}\mathcal{E}}$ to zero also makes the Bures distance independent of time:  $\dB(\rho(t),\rhoD(t)) = \dB(\rho(0),\rhoD(0))$.  This is as expected from the theory of decoherence, because it is the time evolution of the environment states entangled with the system and apparatus that causes decoherence, and with $g_{\mathcal{A}\mathcal{E}}=0$ no such entanglement is generated.

%%%%%%%%%%%%%%%%%%%%%%%%%%%%%%%%%%%
\subsection{Lorentz invariance}
%%%%%%%%%%%%%%%%%%%%%%%%%%%%%%%%%%%
One feature of the model we study is the fact that the Schwinger system is (in the continuum limit) Lorentz invariant. An intriguing question that arises in thinking about decoherence and the ``splitting'' of the wave function into classical branches is the question of when and where in spacetime these splits occur.  To investigate this question, we  compute a mixed state that is localized at site $x$ in the Schwinger space by tracing over the environment, the apparatus, and all the states associated with the Schwinger lattice sites except $x$:
\begin{equation*}
\rho_x \equiv \Tr_{\mathcal{A}, \mathcal{S}\,\text{sites}\, y \neq x}(\rho) \,, \hspace{1.5cm} \rho_{x,D} \equiv \Tr_{\mathcal{A},\mathcal{S}\,\text{sites}\,y \neq x}(\rhoD) \,.
\end{equation*}
\begin{figure}
    \centering
    \includegraphics[height = 10 cm]{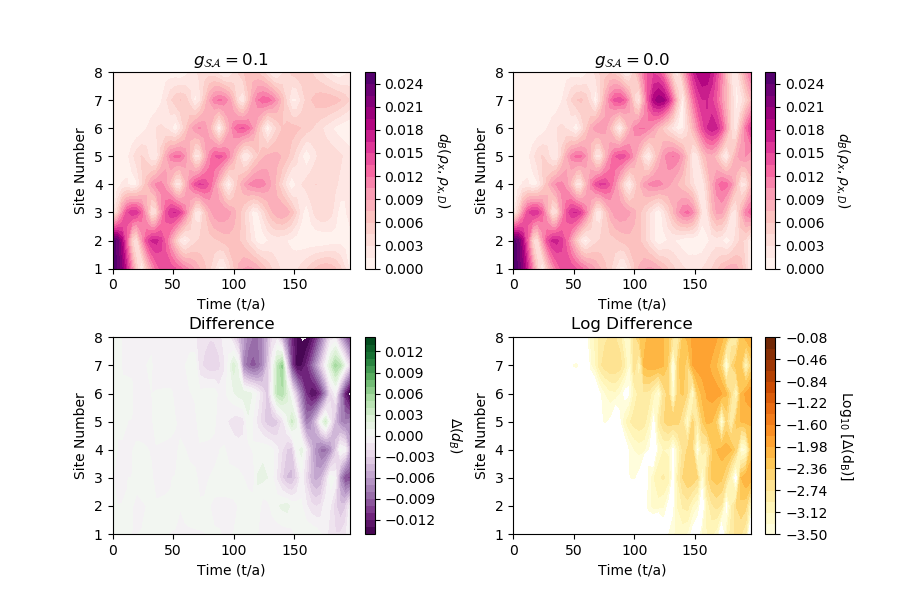} 
    \caption{Top row: Bures distance between the local density matrix $\rho_x$ and the fully decohered local matrix $\rho_{x,D}$, shown for zero and non-zero Schwinger-apparatus coupling. Bottom row: difference and log difference between the two plots in the top row.}
    \label{fig:local}
\end{figure}

The results in this section were obtained with a slightly different interaction Hamiltonian than those in the rest of the paper: the Schwinger operator $\Ctop$ (cf. Eq. \eqref{pointsys}) in this section is the fermion density averaged over only the top two lattice sites of the Schwinger system (as opposed to all but the bottom two).  This makes the interaction more local.

In Figure \ref{fig:local} we plot the Bures distance between $\rho_x$ and $ \rho_{x,D}$.  The difference is initially localized at the site of the charge.  This is as expected: both the states $\ketOmegaS$ and $\ketCS$ are initially very close to the vacuum at the lattice sites away from the position of the charges, and hence are nearly identical. However, as the charges interact with the apparatus (and the apparatus with the environment) we  see that the state $\rho_x$ approaches the decoherered state $\rho_{x,D}$ everywhere.  Intriguingly, this spreading decoherence appears to happen at a speed that exceeds the speed of the charges, but that is slower than the speed of light (which is 1 site/unit time in Figure \ref{fig:local}).  To further investigate this phenomenon will probably require a larger lattice and we leave it for future work.

%%%%%%%%%%%%%%%%%%%%%%%%
\section{Conclusions}
%%%%%%%%%%%%%%%%%%%%%%%%
Our results illustrate that the exact quantum state of our model $\rho$, evolving in time according to the Schr\"odinger equation and with the environment traced over, indeed approaches the  mixed state $\rhoD$ predicted by the theory of decoherence. In the sense that all expectation values are identical,  $\rhoD$ can be regarded as representing a classical statistical ensemble of quantum states with the probabilities predicted to follow a measurement in the conventional interpretations of quantum mechanics.  In this sense our results bear directly on some of the central questions in the interpretation of the quantum wave function and the issue of measurement.

Our study is limited by the computational power of classical computers, which constrains us to consider far smaller Hilbert spaces than those that describe truly macroscopic measuring devices or environments (see e.g.~\cite{Joos:1984uk}). This limitation is likely responsible for the fact that decoherence is only partially effective in our simulations (i.e. for the fact that the  distance between the exact state $\rho$ and the ideal decohered state $\rhoD$ does not decrease to a value very close to zero). As an indication of what would happen with a larger Hilbert space, we investigated both the behavior of the von Neumann entropy of our putative pointer states after some time and the Bures distance between $\rho$ and $\rho_D$ as we increased the size of the apparatus and environment Hilbert spaces. Both results appear to show an increasing level of decoherence.\footnote{Note that extrapolating from $23$ to $10^{23}$ requires a leap of faith!}

%%%%%%%%%%%%%%%%%%%%%%%%
\acknowledgments
%%%%%%%%%%%%%%%%%%%%%%%%
It is a pleasure to thank Aidan Chatwin-Davies, Jonathan Halliwell and Max Schlosshauer for useful comments on a draft of this paper. The work of M.\ K.\ is supported by the NSF through the grant PHY-1820814. O.\ J.\ is supported by the Carl P. Feinberg Graduate Fellowship.

%%%%%%%%%%%%%%%%%%%%%%%%
\bibliographystyle{JHEP}
\bibliography{refs}
%%%%%%%%%%%%%%%%%%%%%%%%
\end{document}